\documentclass[aip,pop,reprint]{revtex4-1}
\usepackage{graphicx}%
\usepackage{dcolumn}%
\usepackage{bm}%
\usepackage{amsmath}%
\draft 

\begin{document}


\title{Creating and studying ion acoustic waves in ultracold neutral plasmas} 



\author{T.C. Killian}
\affiliation{Rice University, Department of Physics and Astronomy and Rice Quantum Institute, Houston, Texas 77005}

\author{P. McQuillen}
\affiliation{Rice University, Department of Physics and Astronomy and Rice Quantum Institute, Houston, Texas 77005}

\author{T.M. O'Neil}
\affiliation{Department of Physics, University of California at San Diego, La Jolla, California 92093}

\author{J. Castro}
\affiliation{Rice University, Department of Physics and Astronomy and Rice Quantum Institute, Houston, Texas 77005}

\date{\today}

\begin{abstract}
We excite ion acoustic waves in ultracold neutral plasmas by imprinting density modulations during plasma creation.  Laser-induced fluorescence is used to observe the  density and velocity perturbations created by the waves. The effect of expansion of the plasma on the  evolution of the wave amplitude is described by treating the wave action as an adiabatic invariant. After accounting for this effect, we determine that the waves are weakly damped, but the damping is  significantly faster than expected for Landau damping.
\end{abstract}


\maketitle 

\section{Introduction}
Collective modes in plasmas \cite{sti92}
are
important dynamical excitations and
 can often be used to obtain information on plasma density, pressure and temperature. Here, we describe the excitation and study of ion acoustic waves (IAWs) in ultracold neutral plasmas (UNPs) \cite{cmk10,kil07,kpp07}.

UNPs are created by photoionizing laser-cooled atoms at the ionization threshold. They have ion and electron temperatures that are orders of magnitude colder than traditional neutral plasmas, so they represent a new regime in which to study   collective effects, where the ions display correlated particle dynamics reflecting strong coupling \cite{ich04,mur06PRL,mth07}. Furthermore, the density distribution and temperatures can be precisely controlled and probed to enable a broad range of experiments.

Experiments and theory on collective modes in UNPs have explored Langmuir oscillations \cite{tla29} excited by RF electric fields \cite{kkb00}. Newly applied techniques of rf absorption \cite{tr11} have confirmed the relationship of the Langmuir oscillations to edge modes and their dependence on non-neutrality induced by plasma dynamics \cite{lpr10,bsp03}. Similar experiments identified Tonks-Dattner modes in a series of RF resonances observed at frequencies above the Langmuir oscillation \cite{fzr06}.
A high-frequency electron drift instability was observed in an UNP in the presence of crossed electric and magnetic fields \cite{zfr08}. Numerical simulations have identified the possibility of propagating spherically symmetric ion density waves \cite{rha03}. A theoretical treatment of IAWs in UNPs, including the effects of strong coupling, was published \cite{shu10} and ion-acoustic shock waves were recently discussed \cite{smd11}. The first experimental excitation of IAWs in UNPs was recently reported \cite{cmk10}. This represents the first study of ion wave motion in UNPs.

 To excite IAWs, we create density perturbations by spatially modulating the intensity of the laser that photoionizes the atoms to create the plasma \cite{cmk10}. Doppler-sensitive laser-induced fluorescence \cite{lcg07,cgk08} is used to study the dispersion relation and damping of the waves and to measure the velocity distribution of the ions.
These techniques  open  new areas of plasma dynamics for experimental study,  including the effects of strong coupling on dispersion relations \cite{rka97,mur98,kaw01,oha00PRL} and non-linear phenomena \cite{nbs99,yra74,rha03,kpp07}
in the ultracold regime.


Low frequency ion density waves in the absence of a magnetic field  are described by the familiar dispersion relation  for frequency
$\omega$ and wavevector $k$ \cite{sti92}
\begin{equation}
\label{eq:dispersionlong}
\left(\frac{\omega}{k}\right)^2=\frac{k_BT_e/m_{i}}{1+k^2\lambda_D^2}
\end{equation}
where $m_{i}$ is the ion mass, $T_e$ is electron temperature,
and $\lambda_D\equiv\sqrt{\epsilon_0k_BT_e/n_ee^2}$ is the Debye screening length, for electron density $n_e$  and charge $e$. We have neglected an ion pressure term because the ion temperature satisfies $T_i\ll T_e$ in UNPs.
In the long wavelength limit, which is the focus here, this mode takes the form of an IAW with $\omega=k\sqrt{k_BT_e/m_{i}}$, in which ions provide the inertia and electron Debye screening moderates the ion-ion Coulomb repulsion that   provides the restoring pressure.

IAWs are highly Landau damped unless $T_i\ll T_e$ \cite{sti92},
however, they are often observable in high-temperature laboratory plasmas \cite{rda60,yra74,nbs99}. Acoustic waves of highly charged dust particles, which show similar characteristics,  have been studied experimentally \cite{bmd95,pgo96} and theoretically \cite{rka97,mur98,oha00PRL,kaw01}. Beyond fundamental interest, IAWs are invoked to explain wave characteristics observed in Earth's ionosphere \cite{koe02} and transport in the solar wind, corona, and chromosphere \cite{cbe07}.

\section{Experimental details}
\subsection{Creation and initial dynamics of an ultracold neutral plasma}
Ultracold neutral plasmas are created by photoionization of laser-cooled strontium atoms in a magneto-optical trap (MOT) \cite{mvs99,scg04,nsl03}. Operating on the dipole-allowed $^{1}\textrm{S}_{0}-^{1}\textrm{P}_{1}$ transition of $^{88}$Sr at $461$ nm, we routinely trap $2\times10^{8}$ atoms at a temperature of $\sim10$ mK in a spherically symmetric Gaussian density distribution, $n(\mathbf{r})=n_{atoms}{\rm exp}(-\mathbf{r}^{2}/2\sigma^{2})$, with $\sigma\approx0.6$ mm and $n_{atoms}\approx6\times10^{10}$ cm$^{-3}$.
Prior to photoionization, the MOT lasers and magnetic field are turned off, and the atoms are allowed to expand  to obtain larger samples, with corresponding lower densities.

Photoionization of the atoms is a two-photon process performed by two tempo-spatially overlapping, $\sim$10 ns laser pulses: the first from a pulse amplified   cw beam tuned to the cooling transition of $461$ nm and the second one from a pulsed dye laser tuned just above the ionization continuum at $\sim412$ nm. The pump for both beams is a 10 ns Nd:YAG laser, operating at 355 nm with a 10 Hz repetition cycle. This process ionizes $\sim$30-70\% of the atoms, creating an ultracold neutral plasma. For a spatially uniform photoionization laser intensity, the plasma inherits its density distribution from the MOT, resulting in initial electron and ion densities ($n_{0e}\approx n_{0i}\equiv n_0$) as high as  $\approx4.2\times10^{10}$ cm$^{-3}$. Effects of un-ionized atoms on the plasma are not significant considering the fast time-scales of the experiment and small neutral-ion collision cross-sections.

Due to their relatively small mass the electrons acquire most of the excess energy from the photoionizing beam, while the ions' kinetic energy remains similar to that of the neutral atoms in the MOT \cite{kpp07}. By tuning dye laser wavelength, we create electrons with initial kinetic  between $1$ and $1000$ K, with a resolution of about 1\,K.

\begin{figure}
\begin{centering}
\includegraphics[width=3in]{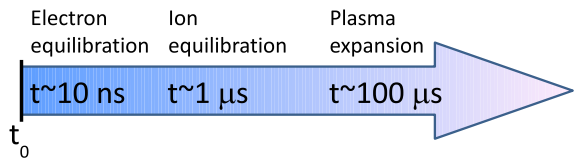}
\par\end{centering}
\caption{Overview of UNP dynamics. Note the three distinct time scales of electron equilibrium, ion equilibrium and global plasma dynamics. (adapted from Ref. \citenum{kpp07}).
\label{fig:TimescalesUNP}}
\end{figure}

On a timescale of the inverse electron plasma oscillation frequency ($1/\omega_{pe}$), electrons equilibrate to produce a nearly thermal distribution \cite{kpp07} (Fig.\ \ref{fig:TimescalesUNP}), yielding average Debye lengths, $\lambda_{D}$ from 3-30\,$\mu$m.  Three-body recombination can also produce large numbers of Rydberg atoms during this initial stage if the electron temperature is low enough \cite{gls07}. (We avoid this recombination-dominated regime for our study of IAWs.)
On a timescale of the inverse ion plasma oscillation frequency ($1/\omega_{pi}$), disorder-induced heating \cite{mur01,csl04} raises the temperature of the cold ions to approximately 1 K.
On a longer hydrodynamic timescale, given by the ratio of the initial characteristic plasma size   to thermal velocities,
\begin{equation}
\tau_{exp}\equiv\sqrt{\frac{m_{{\rm i}}\sigma(0)^{2}}{k_{{\rm B}}[T_{e}(0)+T_{i}(0)]}},\label{eq:texp}
\end{equation}
the plasma expands into the surrounding vacuum \cite{lgs07}. For a strontium plasma with a typical initial size of $\sigma(0)=1$ mm, an initial electron temperature, $T_{e}(0)=50$ K, and an initial ion temperature, $T_{i}(0)=1$ K, $\tau_{exp}= 14$\,$\mu$s.

The expansion of the plasma has a large effect on IAWs that must be accounted for in any quantitative model. Expansion of an unperturbed UNP was studied theoretically in \cite{rha03,mur01,rha02,ppr04PRA,ppr04} and was seen experimentally in \cite{scg04,cdd05,gls07,kkb00,cdd05physplasmas,lgs07,kcg05}. Fundamentally the expansion is
driven by the pressure of the electrons coupled to the ions through a space charge field.


Under the assumptions of quasi-neutrality, adiabaticity, and spherical Gaussian symmetry, the evolution of the size of the plasma cloud and the hydrodynamic expansion velocity ($\mathbf{u}_{exp}$) can be written as
\begin{eqnarray}
\sigma^{2}(t) & = & \sigma^{2}(0)(1+\frac{t^{2}}{\tau_{exp}^{2}}),\label{eq:expansionsigma}\\
\gamma(t) & \equiv & \frac{t/\tau_{exp}^{2}}{1+t^{2}/\tau_{exp}^{2}},\label{eq:expgamma}\\
\mathbf{u}_{exp}(\mathbf{r},t) & = & \gamma(t)\mathbf{r},\label{eq:expvel}
\end{eqnarray}
where $\mathbf{r}$ is the radial distance from cloud center. The ion and electron temperatures ($T_i$ and $T_e$) drop during the expansion due to adiabatic cooling as thermal energy is converted into the kinetic energy of the expansion,
\begin{eqnarray}\label{eq:adiabcool}
T_{\alpha}(t) & = & \frac{T_{\alpha}(0)}{1+t^{2}/\tau_{exp}^{2}}.
\end{eqnarray}
This expansion is similar to dynamics seen in plasmas produced with solid targets, foils, and clusters \cite{lgs07,gpp66,ssc87,bku98,dse98,kby03}.

\subsection{Optical diagnostics of density and velocity distributions}
\label{Sec:opticalprobe}

\begin{figure}
\begin{centering}
\includegraphics[width=2.5in,angle=0 ]{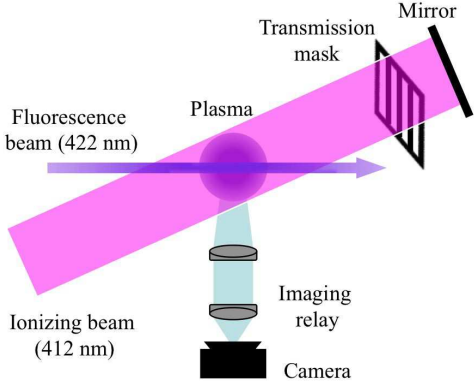}
\par\end{centering}
\caption{Partial experimental schematic showing transmission mask along the path of the ionizing beam.  The ionizing beam is allowed to first pass through the plasma and the transmission mask is placed as close as possible to a retro-reflecting mirror. Due to optical access constraints the k-vector of the IAW is at a $16^\circ$ angle with the k-vector of the fluorescence beam. The fluorescence beam is a sheet perpendicular to the imaging axis (adapted from Ref. \citenum{mck11}).
\label{fig:SculptingDensity}}
\end{figure}

The optical diagnostics used to study UNPs have been described in detail previously \cite{lgs07,cgk08}. The primary probe is Doppler-sensitive laser induced fluorescence (LIF), excited by linearly polarized light of frequency $\nu$ near resonance, $\nu_0$, with the primary $^{2}\textrm{S}_{1/2}-^{2}\textrm{P}_{1/2}$ transition of the Sr ions at $\lambda=$422\,nm. The natural linewidth of this transition is $\gamma_0/2\pi=20$\,MHz. We define $\hat{x}$ as coaxial with the LIF beam, and fluorescence is imaged along the $z$ axis onto  an intensified ccd camera (Figure \ref{fig:SculptingDensity}). To minimize density variations along the line of sight, the excitation light is aligned to the center of the plasma and formed into a sheet in the $x-y$ plane with $1/e^2$ intensity radius $w_z=0.625$\,mm. We derive  spatio-temporally resolved  density and velocity distributions for the ions from the resulting images, $F(x,y,\nu)$. Frequency dependence arises from the natural linewidth, $\gamma_0$, and Doppler-broadening of the transition, and images can be taken with time-resolution of 50\,ns.

A single fluorescence image can be related to underlying physical parameters through
\begin{eqnarray}
F(\nu,x,y) & = & C\frac{\gamma_{0}}{2}\int\hspace{0cm}dz\,n_{i}(\mathbf{r})\frac{I(\mathbf{r})}{I_{sat}} \nonumber \\
&& \times \int d^3\mathrm{v}\,\frac{f(\mathbf{v}){\gamma_{0}}/{\gamma_{eff}}}{1+\frac{I(\mathbf{r})}{I_{sat}}+\left[\frac{2(\nu-\nu_0-\mathrm{v}_x/\lambda)}{\gamma_{eff}/2\pi}\right]^{2}},
 \label{eq:fluorescence}
\end{eqnarray}
where
the multiplicative factor, $C$, depends upon collection solid angle, dipole radiation pattern orientation, and detector efficiency; it is calibrated using absorption imaging \cite{lgs07,cgk08,scg04}.
$I(\mathbf{r})$ is the intensity profile of the fluorescence excitation beam, and $I_{sat}$ is the saturation  intensity for linearly polarized light. Taking Clebsch-Gordon coefficients for the transition into account, $I_{sat}=114$\,mW/cm$^{2}$. $\gamma_{eff}$ is the sum of the natural linewidth and instrumental linewidth. The detuning factor in Eq.\ \ref{eq:fluorescence} reflects the Doppler shift ($\mathrm{v}_x/\lambda$) of the laser frequency for an ion with velocity along the laser of $\mathrm{v}_x$.
The velocity distribution function for the ions, assuming uniform ion temperature, is
\begin{equation}\label{eqn:velocity distribution}
f(\mathbf{v})=\frac{1}{({2\pi}s_{v})^{3/2}}{\rm {exp}}\left\{ -\frac{\left|\mathbf{v}-\mathbf{u}(\mathbf{r})\right|^{2}}{2s_{v}^{2}}\right\},
\end{equation}
with velocity width $s_{v}= \sqrt{k_{B}T_{i}/m_{i}}$.
We allow for a general hydrodynamic velocity $\mathbf{u}$ that can vary with position and will contain expansion velocity (Eq.\ \ref{eq:expvel}) and any other ion motion, such as motion induced by an IAW.

\begin{figure}
\begin{centering}
\includegraphics[clip,width=3in]{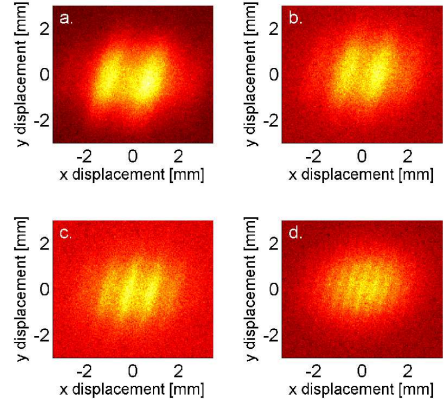}
\par\end{centering}
\caption{ 2D density profiles of density perturbations in UNPs for $T_{e}(0)=48$\,K. The mask wavelengths are (a) 2 mm, (b) 1.33 mm,(c) 1.0 mm, and (d) 0.5 mm.
\label{fig:TwoDIAWimages}}
\end{figure}

A measure of the ion density is obtained by summing the fluorescence signal for a series of images taken at equally spaced frequencies covering the entire ion resonance:
\begin{equation}
\int d\nu\, F(\nu,x,y)=C\frac{\gamma_{0}^{2}}{8}\int dz\,n_{i}(\mathbf{r})\frac{I(\mathbf{r})}{I_{sat}}.\label{eqn:fluorescenceimage}
\end{equation}
For the imaging-laser sheet geometry used here,
$\omega_z\ll\sigma$,   one obtains a measurement of the ion density in the plane of the laser,
\begin{eqnarray}
\,n(x,y,z=0)\approx\frac{\int d\nu\,F(\nu,x,y)}{\left( C\frac{\gamma_{0}^{2}}{8}\frac{I(0)}{I_{sat}}\sqrt{\frac{\pi}{2}}w_z \right)}.
\label{eqn:fluorescenceimagebiglaser}
\end{eqnarray}
Figure \ref{fig:TwoDIAWimages} shows  examples of  two-dimensional false color plots of the ion density distribution.

To perform spatially resolved spectroscopy, which provides information on the ion velocity distribution, we spatially integrate the fluorescence signal given by Eq. \ref{eq:fluorescence} over the region of interest,
\begin{eqnarray}
S_{reg}(\nu) & = & \iint\limits _{region}dxdy\,F(\nu,x,y)\\.
 \end{eqnarray}
A particularly useful expression is found if the region extends over the entire $x-y$ plane to collect signal from all ions illuminated by the laser sheet
\begin{eqnarray}
S_{plasma}(\nu)&\propto& \int d^3\mathrm{{r}}\,\frac{I(\mathbf{r})}{I_{sat}}n_{i}(\mathbf{r}) \nonumber\\
 &&\times \int d\mathrm{v_x}\,\frac{{\gamma_{0}}/{\gamma_{eff}}}{1+\frac{I(\mathbf{r})}{I_{sat}}+\left[\frac{2(\nu-\nu_0-\mathrm{v}_x/\lambda)}{\gamma_{eff}/2\pi}\right]^{2}} \nonumber\\
&&\times \frac{1}{({2\pi}s_{v})^{1/2}}{\rm {exp}}\left\{ -\frac{\left({\mathrm{v}_x}-{\mathrm{u}_x}(\mathbf{r})\right)^{2}}{2s_{v}^{2}}\right\},
\label{eq:fluorsheetspecwholeplasma}
\end{eqnarray}
where $\mathrm{v}_x$ and $\mathrm{u}_x$ are the $x$-components of the velocities, and we have evaluated integrals over $\mathrm{v}_y$ and $\mathrm{v}_z$.
If the hydrodynamic velocity arises purely from self-similar expansion (Eq.\ \ref{eq:expvel}), the spectrum reduces to
a Voigt profile \cite{lgs07,sampadphd},
\begin{eqnarray}
S_{plasma}(\nu)&\propto& \int d\mathrm{v_x}\,\frac{{\gamma_{0}}/{\gamma_{eff}}}{1+\frac{I(\mathbf{r})}{I_{sat}}+\left[\frac{2(\nu-\nu_0-\mathrm{v}_x/\lambda)}{\gamma_{eff}/2\pi}\right]^{2}} \nonumber\\
&&\times \frac{1}{({2\pi}s_{v})^{1/2}}{\rm {exp}}\left\{ -\frac{{v_x}^{2}}{2s_{v,exp}^{2}}\right\}.
\label{eq:fluorsheetspecwholeplasmavoight}
\end{eqnarray}
The rms
width of the Gaussian component of this profile arising
from Doppler broadening reflects
both thermal ion motion
and directed expansion and is given by $s_{v,exp}^2=k_B T_i+\gamma^2\sigma^2$, which increases with time because of plasma expansion (Eq.\ \ref{eq:expvel}).

For a general hydrodynamic velocity, the spectrum may not explicitly take the form of a Voigt profile, but one can show that the width of the spectrum can be quantitatively related to the Lorentzian width, $\gamma_{eff}$, and the mean square x-component of the velocity for all ions in the excitation volume,
$\overline{\left\langle v_x^2\right\rangle}=s_v^2/\lambda^2+   \overline{u_x^2} /\lambda^2$.
\begin{eqnarray} \label{eq:spectralwidth}
  \langle \nu^2 \rangle&=& \int d\nu\,\nu^2S_{plasma}(\nu)/\int d\nu\,S_{plasma}(\nu) \nonumber \\
   &=& 4\alpha \gamma_{eff}^2+\overline{\left\langle v_x^2\right\rangle}/\lambda^2,
\end{eqnarray}
where the overbar indicates  a spatial average and the
mean square x-component of the hydrodynamic velocity is
 \begin{equation}\label{eq:meansquare}
   \overline{u_x^2}=\int d^3\mathrm{{r}}\,\frac{I(\mathbf{r})}{I_{sat}}n_{i}(\mathbf{r})u_x(\mathbf{r})^2.
\end{equation}
The rms width of a Lorentzian is not well-defined, so the factor $\alpha$ depends upon the integration limits in Eq.\ \ref{eq:spectralwidth}. But $\alpha$ is constant and on the order of unity in our experiment, and the Lorentzian contribution to the linewidth is small compared to Doppler broadening, so any error introduced by this effect is small. When the hydrodynamic velocity deviates only slightly from the expansion velocity, which is the case for small amplitude IAWs excited in this study, fitting the spectrum to a Voigt profile provides good approximations of both the Lorentzian width and the the mean square x-component of the total hydrodynamic velocity.

\subsection{Creating density modulations to excite ion acoustic waves\label{sec:sculpdensity}}

In the plasma creation process, uniform intensities of both photoionizing beams, over the length scale of the MOT, result in plasmas with the same density profile as the MOT. However, we can spatially modulate the intensity of either ionizing laser and tailor the initial plasma density distribution. For these studies we have modulated the intensity of the second ionizing beam (412 nm) by placing a transmission mask along its path (Figure \ref{fig:SculptingDensity}).
If we place the transmission mask before a single pass of the ionizing beam through the experimental region, we  create high amplitude modulations in the plasma that access non-linear effects. For this study, however, we probe the linear regime by allowing the 412 nm ionizing beam to first pass through the atoms, then through a transmission mask and back onto the plasma. This  creates $\sim10\%$ plasma density modulation with wavelength set by the period ($\lambda_{0}$) of the mask. We translate the mask pattern to align a density minimum to the center of the plasma, and for this study, we used one-dimensional, 50\% duty cycle, binary patterns with wavelengths of 0.50, 1.00, 1.33 or 2.00 mm (spatial frequencies of 2, 1, 0.75, and 0.5 cy/mm).

\section{Raw data showing IAWs}

Figure \ref{fig:TwoDIAWimages} shows surface plots of the plasma density created with various periodic square wave masks and an initial electron temperature of $T_{e}(0)=48$\,K. We can clearly notice the modulations in density. Due to optical access limitations, the perturbations appear rotated with respect to the field of view of the camera (FOV), i.e. the ionizing beam and fluorescence beam intersect at a $74^\circ$ angle in the imaging plane, so the imaging beam is $16^\circ$ from normal to the IAW propagation axis.

\begin{figure}
\begin{centering}
\includegraphics[clip,width=3in]{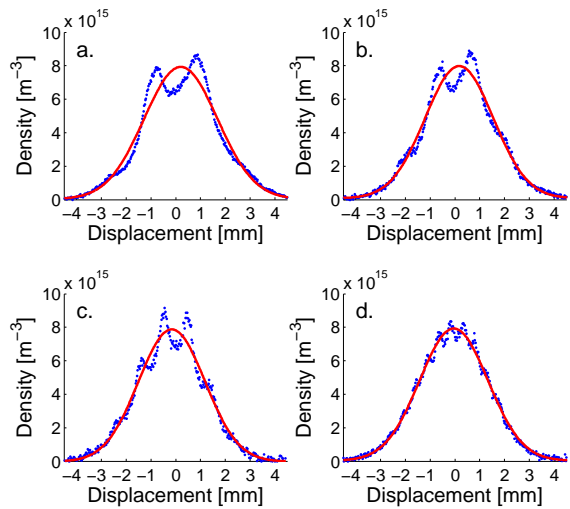}
\par\end{centering}
\caption{1D slices through density profiles for $T_{e}(0)=48$\,K showing
density perturbations.
 The mask wavelengths are (a) 2 mm, (b) 1.33 mm,(c) 1.0 mm, and (d) 0.5 mm. Dots are data and deviations from the fit Gaussian (solid line) represent the IAW density modulation, $\delta n$ (Eq.\ \ref{eq: totaldensity}).
\label{fig:GaussPerturb}}
\end{figure}

\begin{figure}
\begin{centering}
\includegraphics[width=3in]{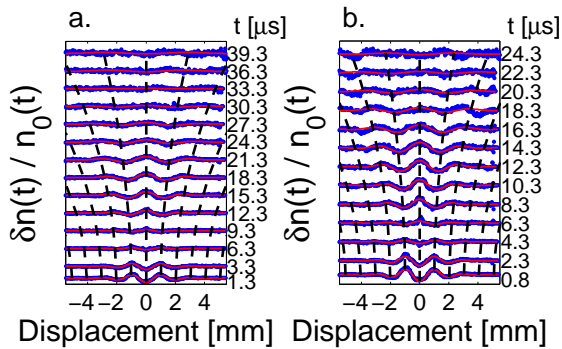}
\par\end{centering}
\caption{Evolution of $\delta n$ for mask wavelength of 2 mm. Time since ionization is indicated on the right, and $\delta n(x,t)$ has been scaled by instantaneous peak density $n_{Gauss}(0,t)$ and offset for clarity, for (a) $T_{e}(0)=48$\,K, (b) $T_{e}(0)=105$\,K. Solid red lines are fits to Eq. (\ref{eq:wavemodel}), and the dotted black lines follow points of constant spatial phase of the standing wave.  As the electron temperature increases, the frequency of the wave increases and for a fixed electron temperature, as the number of cycles per millimeter increases, the frequency of the wave also increases.
\label{fig:EvolPerturb05cymm}
}
\end{figure}

To study the evolution of the density modulations, we form 1D density profiles from a central strip of width $\sigma$ by averaging the 2D data  parallel to the modulations (Fig.\ \ref{fig:GaussPerturb}). By fitting such data to a Gaussian, we can separate the total density into background $n_{Gauss}$ and wave $\delta n$ components
\begin{equation} \label{eq: totaldensity}
n_i(x,t)=n_{Gauss}(x,t)+\delta n(x,t).
\end{equation}
 Figure \ref{fig:EvolPerturb05cymm} shows  the evolution of $\delta{n(x,t)}$ for a one transmission mask period with different initial electron temperatures. Note the oscillation of the wave in time and space.

\section{Effect of plasma expansion on the IAW\label{sec:plasmaexpansion}}
To motivate the formalism for analyzing IAWs, it is necessary to understand the effect of the plasma expansion on the IAW amplitude. This can be found by treating the wave action as an adiabatic invariant \cite{whi65}. We start by describing an ion-acoustic wave of wave vector $k$ and amplitude $\delta n$ in an infinite homogeneous plasma, for which the dielectric function is given by
\begin{equation}\label{Eq:dielectricfunction}
    \varepsilon(k,\omega)=1-\frac{\omega^2_{pi}}{\omega^2}+\frac{1}{k^2\lambda^2_{D}}.
\end{equation}
This implies that the wave energy density is
\begin{equation}\label{Eq:waveenergydensity}
    W_k=\frac{\partial}{\partial \omega}\left[\omega \varepsilon(k,\omega)
    \right ]\frac{|E_k|^2}{8\pi}= \frac{\omega^2_{pi}}{\omega^2}\frac{|E_k|^2}{8\pi}
\end{equation}
for wave field amplitude $E_k$, $\omega_{pi}\gg \omega$, and $2\pi/k\ll \lambda_{D}$. The field amplitude can be related to the amplitude of the IAW through
\begin{equation}\label{Eq:fieldandIAWamplitude}
  |\delta n|=  n_i\frac{e}{m_i} \frac{k}{\omega^2}|E_k|.
\end{equation}
The action density is given by ${W_k}/{\omega}$, which yields the total action in a volume $V$
\begin{equation}\label{Eq:action}
   N_k=  \frac{|\delta n|^2 m_i \omega}{2 n_i k^2}V.
\end{equation}

For a slow expansion ($|\dot{\omega}/\omega^2| \ll 1$) and if damping effects such as Landau damping are negligible,  the IAW will evolve adiabatically during the expansion, and  the total IAW action will remain constant. For our measurements, $|\dot{\omega}/\omega^2 | \ll 1$ is a good approximation at early times, although for longer wavelengths and long times, $|\dot{\omega}/\omega^2 |\approx 1$.

 In our inhomogeneous UNPs, the length scale for density variations is large compared to the characteristic scale of the IAW ($\sigma \gg 2\pi/k$), which suggests  that Eq.\ \ref{Eq:action} should apply locally \cite{hto10}. This allows us to identify $V(t)\propto \sigma(t)^3$ and $n_i\propto 1/\sigma^3(t)$. As shown previously \cite{cmk10} and discussed below, $k(t)\propto 1/\sigma(t)$ and $\omega(t) \approx k(t)\sqrt{T_e(t)/m_i}\propto 1/\sigma(t)^2$, where we have used the time evolution of the electron temperature (Eq.\ \ref{eq:adiabcool}). For constant $N_k$, this yields the final result for the evolution of the IAW amplitude $|\delta n(x,t)|\propto 1/\sigma^3(t)\propto n_{Gauss}(0,t)$. So we expect the wave amplitude to scale with the the peak ion density, $n_{Gausss}(0,t)$, and deviation from this behavior would indicate damping or gain processes.

\section{Measuring IAW wavelength, frequency, and dispersion}
To analyze data for IAW evolution such as to measure the wavelength, frequency, dispersion, and damping rates, we fit the data to a standing wave model.
As discussed in the previous section,  we scale the amplitude of the perturbations by the instantaneous peak density  obtained from the Gaussian fit (Fig. \ref{fig:GaussPerturb}), and assume a damped standing wave with a Gaussian envelope,
\begin{eqnarray}
\frac{\delta{n(x,t)}}{n_{Gauss}(0,t)}  &=&  \frac{\delta{n(0,t)}}{n_{Gauss}(0,t)}e^{-x^{2}/2\sigma_{env}(t)^{2}}
 \cos{[k(t)x]}\nonumber \\
  &=&  A(t)e^{-x^{2}/2\sigma_{env}(t)^{2}}\cos{[k(t)x]}.\label{eq:wavemodel}
\end{eqnarray}
This expression fits the modulation at a single time, $t$, and yields the instantaneous amplitude $A(t)$,  envelope size $\sigma_{env}(t)$ and wavevector $k(t)$. The hydrodynamic description for an infinite homogeneous medium used to derive the IAW dispersion relation, Eq. \ref{eq:dispersionlong}, allows for planar standing-wave solutions, but finite size, plasma expansion, and density inhomogeneity introduce additional factors that are not small here and preclude an analytic solution, so the model behind Eq. \ref{eq:wavemodel} is only phenomenological. The solid red lines in Figure \ref{fig:EvolPerturb05cymm} are fits to this evolution model in which $A$, $\sigma_{env}$, and $k$ are allowed to vary independently for each curve.

The dotted black lines in Figure \ref{fig:EvolPerturb05cymm} follow trajectories of constant $k(t)x$. Notice that as the wave evolves, the wavelength/wavevector appears to increase/decrease with time. If we hypothesize that the wavelength scales with plasma expansion (Eq.\ \ref{eq:expansionsigma}), then we have:
\begin{equation}
k(t)=\frac{k_{0}}{\left({1+t^{2}/\tau_{exp}^{2}}\right)^{1/2}}.
\label{eq:KVecEvol}
\end{equation}
Fits to $k(t)$ measurements, with $k_0$ and $\tau_{exp}$ as fit parameters, yield values for $2\pi/k_{fit,0}$ that match the transmission mask wavelength. We also find very good agreement between the extracted fit value for $\tau_{exp}$ and the value expected from self-similar expansion, Eq. \ref{eq:texp}.

\begin{figure}
\begin{centering}
\includegraphics[width=3.in]{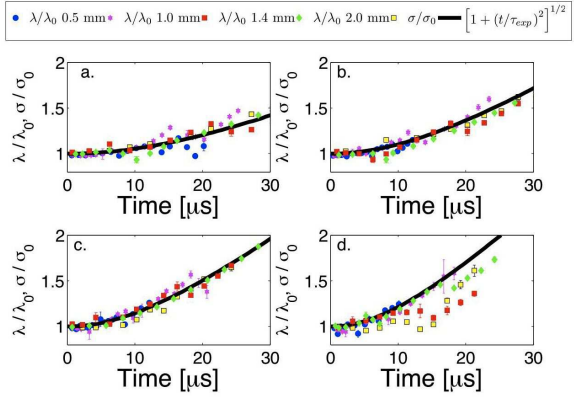}
\par\end{centering}
\caption{Evolution of the IAW wavelength and plasma size, normalized to initial
values, for $\sigma_{0}=1.45$\,mm and for (a) $T_{e}(0)=25$\,K, (b) $T_{e}(0)=48$\,K, (c) $T_{e}(0)=70$\,K, and (d) $T_{e}(0)=105$\,K. Size and wavelength follow a universal curve with no fitting parameters. For the solid line, $\tau_{exp}$ has been set to its theoretical value (Eq. \ref{eq:texp}). The universal behavior for wavelength and plasma size indicate the wave is pinned to the expanding plasma.
\label{fig:KvectorEvolAllTes} }
\end{figure}

To show this behavior, we compare the evolution of the wavelength with the evolution of the size of the plasma by plotting, in Fig. \ref{fig:KvectorEvolAllTes}, $\lambda(t)/\lambda_{0}$, $k_{0}/k(t)$ and $\sigma(t)/\sigma_{0}$, where all quantities have been normalized to the values at $t=0$. Initial wavelengths match the period of the mask used and all the data follow one universal curve, $\left({1+t^{2}/\tau_{exp}^{2}}\right)^{1/2}$, with no fit parameters and $\tau_{exp}$ as calculated from $\sigma_{0}$ and $T_{e}(0)$ (Eq. \ref{eq:texp}). This universal behavior for wavelength and plasma size indicate the wave is pinned to the expanding density distribution.

\begin{figure}
\begin{centering}
\includegraphics[width=3in]{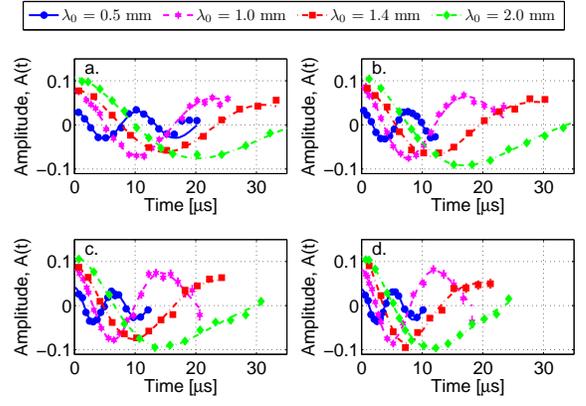}
\end{centering}
\caption{Evolution of the amplitude of
$\delta n(x,t)/n_{Gauss}(0,t)$ (Eq.\ \ref{eq:wavemodel})
and fits to obtain $\omega(t)$ for (a) $T_{e}(0)=25$\,K, (b) $T_{e}(0)=48$\,K, (c) $T_{e}(0)=70$\,K, and (d) $T_{e}(0)=105$\,K. Lines are fits to Eq. (\ref{eq:ampevol}) in which $\tau_{exp}$ has been fixed to the theoretical value.}
\centering{}\label{fig:AmplitudeEvolutionAllTes}
\end{figure}

To obtain a dispersion relation, we now need to extract the frequency of the wave, $\omega(t)$. We can extract the frequency by analyzing the evolution of the perturbation amplitude. The amplitude variation in time can be modeled as
\begin{equation}
A(t)=A_{0}\mathrm{e}^{-\Gamma t}\cos\left[\phi(t)\right]=A_{0}\mathrm{e}^{-\Gamma t}\cos\left(\int_{0}^{t}\omega(t')dt'\right),\label{eq:ampevol}
\end{equation}
 where the integral accounts for accumulation in the phase. By fitting the evolution of the phase of the entire wave, we are essentially finding the frequency in the frame moving with each plasma element, and our measurement does not need to be corrected for Doppler shifts. For the frequency, we now assume the form of the dispersion for an infinite, homogeneous medium from Eq. \ref{eq:dispersionlong}, the observed variation of wave-vector $k(t)$ (Eq. \ref{eq:KVecEvol}), and electron temperature evolution $T_{e}(t)$ predicted for a self similar expansion (Eq. \ref{eq:adiabcool}) to obtain
\begin{eqnarray}
\omega(t)  =  k(t)\sqrt{k_{B}T_{e}(t)/m_{i}}=\omega_{0}\left(\frac{1}{1+t^{2}/\tau_{exp}^{2}}\right).\label{eq:omegaevol}
\end{eqnarray}
 Initial amplitude $A_{0}$, frequency $\omega_{0}$, and damping rate $\Gamma$ are allowed to vary in the fits, while $\tau_{exp}$ is taken as the value predicted for the electron temperature set by the laser and initial plasma size, (Eq. \ref{eq:texp}) resulting in fits that match the data very well (Fig. \ref{fig:AmplitudeEvolutionAllTes}). As can be seen from Fig. \ref{fig:AmplitudeEvolutionAllTes} and Eq. \ref{eq:omegaevol}, the frequency decreases with time. Similar behavior was observed in simulations of spherical IAWs \cite{rha03}.


\begin{figure}
\begin{centering}
\includegraphics[width=3in]{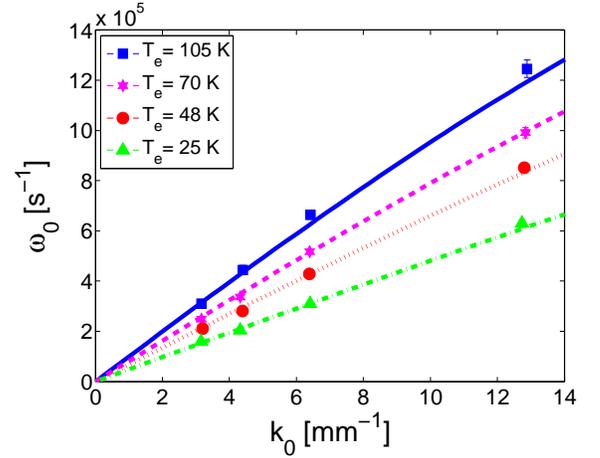}
\par\end{centering}
\caption{{Dispersion relation of IAWs for different initial electron temperatures.} Lines are from the theoretical dispersion relation, Eq. \ref{eq:dispersionlong}, with no fit parameters. $k_{0}\lambda_{D}<0.42$ for all conditions used in this study, with the average density being used to calculate $\lambda_{D}$.}
\centering{}\label{fig:DispersionRelation}
\end{figure}

We extract $\omega_{0}$ and $k_{0}$ for a range of initial electron temperatures and mask periods, and calculate the dispersion of the excitations, as shown in Fig. \ref{fig:DispersionRelation}. The excellent agreement with theory (Eq. \ref{eq:dispersionlong}) confirms that these excitations are IAWs. The planar standing-wave model captures the dominant behavior of the wave, and to a high accuracy there is no deviation from the standard dispersion relation in spite of the plasma's finite size, expansion, and inhomogeneous density. We suspect that this follows from the fact that the length scale for background density variation and time scale for plasma expansion are reasonably long compared to the IAW wavelength and period respectively. Observations for longer times will be more sensitive to interactions between the wave and the boundary.

\begin{figure}
\begin{centering}
\includegraphics[width=3in]{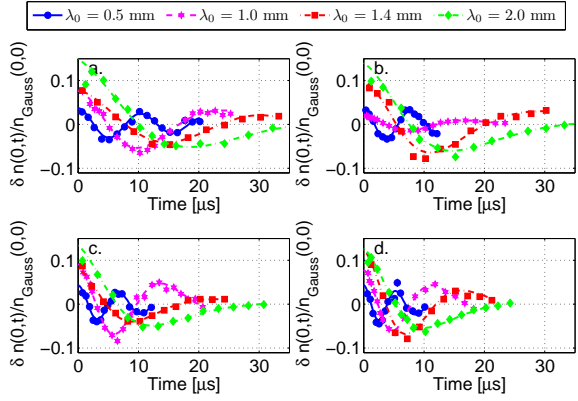}
\end{centering}
\caption{Evolution of
the amplitude of
$\delta n(x,t)/n_{Gauss}(0,0)$
for (a) $T_{e}(0)=25$\,K, (b) $T_{e}(0)=48$\,K, (c) $T_{e}(0)=70$\,K, and (d) $T_{e}(0)=105$\,K, as was done for analysis in the first report of IAWs in UNPs\cite{cmk10}.}
\centering{}\label{fig:AmplitudeEvolutionAllTesOldScaling}
\end{figure}

In the first report of IAWs\cite{cmk10}, the frequency was found by fitting the evolution of $\delta n(x,t)/n_{Gauss}(0,0)$ (Fig.\ \ref{fig:AmplitudeEvolutionAllTesOldScaling}) instead of $\delta n(x,t)/n_{Gauss}(0,t)$ (Eq.\ \ref{eq:wavemodel}). Within our uncertainties,   there is no difference in the resulting dispersion relation.

\section{IAW damping}

The scaling of the amplitude has a significant effect on the interpretation of the damping of IAWs. The unscaled amplitude of the perturbations, $\delta n$, decreases rapidly in time (Fig.\ \ref{fig:AmplitudeEvolutionAllTesOldScaling}), but as discussed in Sec.\ \ref{sec:plasmaexpansion}, a significant decrease is expected due to plasma expansion, which is distinct from damping. This effect is accounted for by scaling the amplitude by the instantaneous peak density, $\delta n(x,t)/n_{Gauss}(0,t)$, which yields $A(t)$ shown in Fig.\ \ref{fig:AmplitudeEvolutionAllTes}.  With proper scaling, it is clear that damping is  small on the timescale of our measurements.

\begin{figure}
\begin{centering}
\includegraphics[width=2.5in, angle=0]{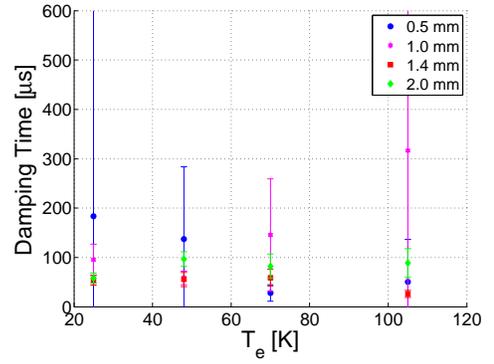}
\par\end{centering}
\caption{Measured damping times ($\Gamma^{-1}$) found from fitting the wave amplitude evolution
to Eq.\ \ref{eq:ampevol}. Initial IAW wavelengths are given in the legend.}
\label{fig:DampingScaled}
\end{figure}

The inverse of the damping rate, found by fitting Eq.\ \ref{eq:ampevol} to the data, is shown in Fig.\  \ref{fig:DampingScaled}. The typical damping times are 3-10 times longer than the timescale of our measurements, and thus poorly determined. The observation times were limited by decreasing density and signal as the plasma expanded. Shorter wavelengths \cite{mck11} will increase the oscillation frequency and allow us to observe more periods and explore the damping with greater precision.

The damping of collective effects is interesting in plasmas because it can be sensitive to collisional or kinetic effects \cite{lan46}. Damping can also probe important many-body properties such as viscosity, which has become a topic of great interest  in strongly-coupled systems \cite{tho09,mth07}.
In the case of IAWs, Landau damping is often the dominant damping mechanism. When $T_e\approx T_i$ there is a large population of electrons traveling at just below the phase velocity of the wave that can efficiently extract energy from the wave.  The Landau damping  rate scales with the oscillation frequency, $\omega$ (Eq.\ \ref{eq:dispersionlong}), as given by \cite{bth10,lan46}
\begin{eqnarray}
\label{eq: landau damping}
\gamma_L & \approx & \omega \frac{\sqrt{\pi/8} }{\left(1+k^2\lambda_D^2\right)^{3/2}} \nonumber\\
&  &\times \left[\left(\frac{m_e}{m_i}\right)^{1/2} + \left(\frac{T_e}{T_i}\right)^{3/2}\mathrm{exp}\left(\frac{-T_e/T_i}{2\left(1+k^2\lambda_D^2\right)}\right)\right].
\end{eqnarray}
To account for the time dependence of the damping rate over the evolution time of the experiment we average the calculated rate over  the  observation time,
\begin{eqnarray}
\label{eq:AverageRate}
\gamma_{avg}\equiv \frac{1}{t_{0}}\int_0^{t_0}\gamma_L \left( t \right) \mathrm{d} t.
\end{eqnarray}
The inverse of this average rate gives  theoretical damping times of larger than 1\,ms for all conditions studied here, which is much longer than the timescale observed for damping of IAWs.

%

 Intuitively, one  expects other effects to lead to damping or dephasing of the wave in UCPs. For example,  propagation into low density regions at the plasma outer boundary should lead to decay.  Also, when the wavelength decreases and the dispersion relation deviates from  constant phase velocity due to the nonvanishing Debye screening length ($k\lambda_D\approx 1$), the dispersion relation should depend on plasma density and vary through the plasma. This will lead to a more complex wave evolution.

It would  be very interesting to access a regime in which Landau damping dominates. It is difficult to achieve significantly lower $T_e/T_i$ in UNPs because of intrinsic electron heating effects at low temperature, such as three body recombination \cite{gls07}. But at short wavelength, as $k\lambda_D $ approaches unity, the damping increases sharply and should be observable. 
Much more work remains to be done to understand damping of IAWs in UNPs, nonetheless, this work represents a valuable first step.

\section{Ion velocities in the IAWs}

In the previous sections, we have studied IAWs through the variation of density in space and time. It is also possible to measure the velocity of the ions in the plasma, which should also display the effects of the oscillation.

For total density given by Eq.\ \ref{eq: totaldensity} with perturbation $\delta n$ given by the standing wave model (Eq.\ \ref{eq:wavemodel}), the ion flux $J$ can be found from the one-dimensional continuity equation, $\partial n/\partial t=-\partial J/\partial x$. For IAW wavelength much less than the characteristic size of the plasma ($\lambda\ll \sigma$) and slow variation of   $\sigma$, $\omega$, and $\lambda$, the flux is

\begin{equation}\label{Eq:IAWflux}
    J\approx\frac{\omega}{k}A_0 n_{Gauss}(0,t) \mathrm{e}^{-x^2/2\sigma_{env}^2}\mathrm{e}^{-\Gamma t}\,\mathrm{sin}kx \,\mathrm{sin}\omega t.
\end{equation}
The hydrodynamic ion velocity at $x$ and $t$ due to an IAW is given by $u_{IAW} ={J}/{n}$. We observe that the envelope size is approximately equal to the characteristic plasma size ($\sigma_{env}\approx \sigma$). So exponential factors in this ratio cancel, and
\begin{equation}\label{Eq:IAWvelocity}
    u_{IAW}
    \approx \frac{\omega}{k}A_0 \mathrm{e}^{-\Gamma t}\,\mathrm{sin}kx \,\mathrm{sin}\omega t,
\end{equation}
where we have assumed a small perturbation. In addition to the IAW velocity, ions possess the hydrodynamic velocity due to plasma expansion (Eq.\ \ref{eq:expvel}) and random velocities characterized by the ion temperature $T_i$.

Section \ref{Sec:opticalprobe} describes how the Doppler-sensitive laser-induced fluorescence spectrum can be used to obtain the mean square of the velocity along the laser propagation direction for all ions in the excitation volume,
$\overline{\left\langle v_x^2\right\rangle}$ (Eq.\ \ref{eq:spectralwidth}), where the overbar indicates  a spatial average.  For a small-amplitude IAW in an expanding plasma, $\overline{\left\langle v_x^2\right\rangle}$ arises from thermal, expansion, and IAW contributions,
\begin{equation}\label{Eq:plasmaaveragekineticenergy}
    \overline{\left\langle v_x^2\right\rangle} = {k_B T_i}/m_i + \gamma^2 \sigma^2+ \overline{ u_{IAW}^2}+\overline{2\gamma x u_{IAW}}.
\end{equation}
The average of the cross term, $\overline{2\gamma x u_{IAW}}$, tends to zero for small IAW wavelength, and is already small for our conditions. We will neglect it in subsequent analysis. For clarity, we have also neglected   effects of the small angular misalignment of the IAW and laser propagation directions.

\begin{figure}[ht]
\begin{centering}
\includegraphics[width=3in,angle=0]{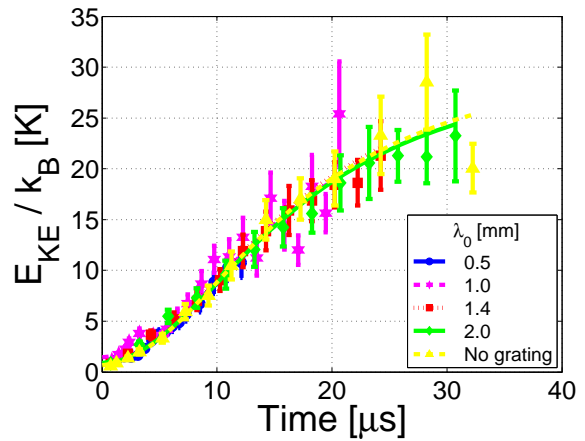}
\par\end{centering}
\caption{Evolution of the average ion kinetic energy for $T_e(0)=70$\,K, and various IAW wavelengths or no IAW. Data are fit to $\frac{1}{2} m_{i}  \overline{\left\langle v_x^2\right\rangle} = \frac{1}{2}{k_B T_i} + \frac{1}{2} m_{i}\gamma^2 \sigma^2$. This emphasizes that the data follows a universal curve indicating a negligible effect of the modulations on the overall expansion.
}
\centering{}\label{fig:ComparisonExpVelocityFitAllTesOverlay}
\end{figure}

Figure \ref{fig:ComparisonExpVelocityFitAllTesOverlay} shows measurements of
$\overline{\left\langle v_x^2\right\rangle}$ extracted from the spectra for $T_e(0)=70$\,K
expressed in terms of an average kinetic energy of the ions, $m_{i}  \overline{\left\langle v_x^2\right\rangle}/2$.
The small offset for the kinetic energy at early times arises from disorder-induced heating of the ions \cite{scg04,csl04}. The kinetic energy then shows a gradual increase on the timescale of $\tau_{exp}$
as electron energy is transferred into expansion velocity of ions. Finally, it approaches a terminal velocity $\sim k_BT_{e0}/2m_{i}$ when all the energy has been transferred. Note that the overall evolution of kinetic energy is affected very little by the presence of IAWs, which further confirms that the plasma expansion is not affected significantly by the small IAWs. The only deviations are small oscillations at early times.

\begin{figure}[ht]
\begin{centering}
\includegraphics[width=3in, clip=true,trim=0 0 00 00,angle=0]{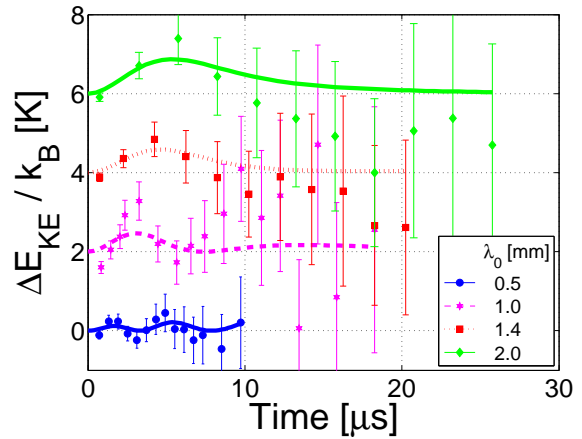}
\par\end{centering}
\caption{Oscillations in the ion kinetic energy for $T_e(0)=70$\,K, extracted from the data shown in Fig. \ref{fig:ComparisonExpVelocityFitAllTesOverlay} obtained upon subtraction of the calculated average ion kinetic energy from the fits shown in Fig. \ref{eq: IAWkineticenergy}. The solid lines represent a fit to Eq. \ref{eq: IAWkineticenergy} and for clarity, the curves are offset.}
\centering{}\label{fig:ComparisonExpVelocityOscAllTes}
\end{figure}

Figure \ref{fig:ComparisonExpVelocityOscAllTes} shows the difference between the data and the fits to
$\frac{1}{2} m_{i}  \overline{\left\langle v_x^2\right\rangle} = \frac{1}{2}{k_B T_i} + \frac{1}{2} m_{i}\gamma^2 \sigma^2$, which emphasizes the effect of the IAW. This difference is now fit to
\begin{equation}\label{eq: IAWkineticenergy}
   \Delta E_{KE}=\frac{1}{2} m_{i} \overline{u_{IAW}^2}= \frac{1}{4} m_{i}\left(\frac{\omega}{k}A_0\mathrm{e}^{-\Gamma t}\right)^2 \,\mathrm{sin}^2\omega t
\end{equation}
 plus an arbitrary offset, where the evolution of $\omega/k$ is replaced by the expected IAW speed $\sqrt{k_B T_e(t)/m_{i}}$, the evolution of $\omega$ is set by Eq.\ \ref{eq:omegaevol}, and the only fit parameters are the fractional amplitude $A_0$ and decay rate $\Gamma$. Notice the clear oscillations in the kinetic energy  at twice $\omega$ that are out of phase with the oscillations in amplitude of the density perturbation shown in Fig.\ \ref{fig:AmplitudeEvolutionAllTes}.

Figure \ref{fig:ComparisonExpVelocityOscAllTes} can be interpreted as a measurement of the amount of energy in the waves. This energy oscillates back and forth between potential energy of ions in the electric field when the density perturbation is largest and kinetic energy when the kinetic energy is largest. Measuring velocity perturbations may prove to be a more valuable probe of IAWs than density perturbations when the wavelength is very small and difficult to resolve optically .

\section{Conclusions}
We have presented expanded studies of ion acoustic waves in ultracold neutral plasmas \cite{cmk10}, including a description of the effects of plasma expansion on the amplitude of the wave. We observe the wave through  density and velocity perturbations, and we find excellent agreement with the well known IAW dispersion relation. The IAWs damp significantly faster than expected for Landau damping, and this effect remains to be explained.  In the future, we plan to extend these measurements to the short wavelength \cite{mck11}, non-linear dispersion regime of ion plasma oscillations, where strong coupling is predicted to have an effect and Landau damping increases.
With these tools we can also study different initial plasma geometries to access other phenomena such as streaming plasmas, nonlinear waves, and possibly shock physics.


%
%

%

\begin{acknowledgments}
Financial support for this work was provided by the Department of Energy and the National Science Foundation.
\end{acknowledgments}


%

\end{document}